\documentclass[usenatbib]{mn2e}
\input epsf
\title[Millisecond dips of Sco X-1 and TNO occultation]{Millisecond dips 
in the RXTE/PCA light curve of Sco X-1 and TNO occultation}
\author[Chang, Liang, Liu, and King]{Hsiang-Kuang Chang$^{1,2}$\thanks{E-mail:
hkchang@phys.nthu.edu.tw},
Jau-Shian Liang$^1$,
Chih-Yuan Liu$^2$,
and 
Sun-Kun King$^3$
\\
$^{1}$Department of Physics, National Tsing Hua University, Hsinchu 30013, Taiwan\\ 
$^{2}$Institute of Astronomy, National Tsing Hua University, Hsinchu 30013, Taiwan\\ 
$^{3}$Institute of Astronomy and Astrophysics, Academia Sinica, Taipei 10617, Taiwan} 

\begin{document}

\date{Accepted 2007 Month date. Received 2007 Month date; in original form 2007 Month date}

\pagerange{\pageref{firstpage}--\pageref{lastpage}} \pubyear{2007}

\maketitle

\label{firstpage}

\begin{abstract}
Millisecond dips in the RXTE/PCA light curve of Sco X-1 were reported recently 
(Chang et al.\ 2006), which were interpreted as the occultation of X-rays from Sco X-1
caused by Trans-Neptunian Objects (TNO) of hundred-meter size.
Inconclusive signatures of possible instrumental effects in many of these dip events 
related to high-energy cosmic rays
were later found (Jones et al.\ 2006) and the TNO interpretation became shaky.
Here we report more detailed analysis aiming at distinguishing 
true occultation events from those related to cosmic rays. 
Based on some indicative criteria derived from housekeeping data and two-channel
spectral information, we suggest that about 10\% of the dips are probable 
events of occultation. The total number of TNOs of size from 60 m to 100 m
is estiamted to be about $10^{15}$ accordingly.
Limited by the coarser time resolution of standard data modes of RXTE/PCA, however, 
definite results cannot be obtained. Adequately configured observations
with RXTE or other new instruments in the future are very much desired.
\end{abstract}


\section{Introduction}

The outer solar system beyond the orbit of Neptune 
is a relatively unfamiliar world 
to the human being, 
and yet it is believed to contain debris of 
the primordial solar disk and therefore carries much information
about the early solar system. 
Although the existence of objects 
in the outer solar system was proposed long time ago 
\citep{leonard30,edgeworth43,edgeworth49,oort50,kuiper51},
except for Pluto discovered by C. W. Tombaugh in 1930,
the first of such objects
was not found until 1992 \citep{jewitt93}.
Since then several surveys have been performed and about 
1000 TNOs have been discovered 
(http://cfa-www.harvard.edu/cfa/ps/lists/TNOs.html). 

The differential size distribution of the observed TNOs, 
which are all larger than about 100 km in diameter, 
follows a power law like
${\rm d}N/{\rm d}s\propto s^{-q}$ with 
$q=4.0\pm 0.5$ (e.g.\ \citet{trujillo01,luu02}).
The size distribution, in addition to
representing a major population property of TNOs, 
provides important constraints on the theory of how our planetary
system was formed. 
In coagulation models of planet formation with 
collisional disruption cascade, for example, 
the power-law size distribution 
does not simply extend towards smaller size but turns flatter 
(a smaller $q$) below a certain break radius 
(e.g.\ \citet{farinella00,kenyon02,kenyon04}). 
The magnitude of the break radius depends on the initial mass 
in the trans-Neptunian region, 
the epoch of Neptune's formation, and the tensile strength 
of these small bodies. Extending our knowledge 
of the size distribution 
to smaller TNOs is clearly desirable.  

TNOs smaller than 100 km are too dim to detect. 
A Hubble Space Telescope survey reported the detections 
of 3 TNOs within 0.02 square degrees
of the sky, whose estimated diameters are about 30 km
\citep{bernstein04}.
These detections were achieved by blindly integrating images 
repeatedly over all possible TNO orbits for a total exposure of 22 ks. 
These 3 detections are a factor of 
25 less than that expected from the extrapolation of the size distrbution 
of TNOs larger than 100 km.
For even smaller TNOs, occultation of background 
stars as a way to study their properties was proposed 
thirty years ago \citep{bailey76,brown97,cooray03,alcock03,roques03}.
Searches for such occultation events in optical bands have 
been conducted by many groups but without 
any definite detection so far. 
Three possible detections were just reported recently \citep{roques06}.

On the other hand, occultation events in X-rays may stand 
a better chance of detection, compared with the optical band. 
The photon-counting nature of X-ray detectors like proportional 
counters provides much faster photometry 
and the much shorter wave length of X-rays means much less diffraction. 
These two features make it more feasible to
detect events caused by smaller bodies, 
and henceforth can provide a larger number of possible events. 
This technique requires a bright background X-ray source. 
Sco X-1, the brightest X-ray source in the sky, 
serves this purpose well.
Its location, only about 5.5 degrees north of the ecliptic,
allows good sampling of the TNO population.
The Proportional Counter Array (PCA) on board Rossi X-ray
Timing Explorer (RXTE) \citep{bradt93},
a NASA satellite, has the largest effective area in the keV X-ray band. 
The typical RXTE/PCA count rate of Sco X-1 is $10^5$ counts per second. 
It is therefore possible to search 
for occultation events
at millisecond time scales. 

Millisecond dip events in the X-ray light curve of Sco X-1, in total 58 in 322-ks data,
were reported recently 
\citep{chang06}, which were attributed to occultation caused by small TNOs.
Later, signatures of possible instrumental effects related to high energy cosmic rays
for many of these dip events were found \citep{jones06} and their nature 
became uncertain. 
In this paper, we report further analysis of 107 dip events (including  
the original 58), which were 
found 
in 564-ks RXTE archival data spanning over 7 years from 1996 to 2002.
The procedure to identify these events and their properties are described in Section 2.
Signatures of possible instrumental effects are discussed in Section 3.
In Section 4 we propose some criteria to tentatively separate true 
astronomical events from those likely due 
to the instrumental dead-time effect.  
We then further argue that those dip events 
without indication of instrumental effects
are occultation caused by small TNOs and derive the total number of TNOs 
in the corresponding size range in Section 5.   

\section{Millisecond dips in the RXTE/PCA light curve of Sco X-1}

To search for possible occultation events, we examined the light curves of Sco X-1 
binned in different bin sizes
from all the available RXTE/PCA archival data
and computed their deviation distributions. 
%
The definition of  the `deviation' employed here can be found in \citet{chang06}.
The deviation distribution of all the time bins in the light curve 
should be of a Poisson nature modified by the dead-time and coincidence-event effects 
if all the fluctuations are random 
and there is no occultation event. 
One example of such a deviation distribution is shown in Fig. 1, 
in which the search was 
conducted with 2-ms time bins. The excess at the negative deviation end is obvious. 
We performed the search with different time-bin
sizes ranging from 0.5 ms to 20 ms. In each search, those time bins with a negative deviation 
of random probability lower than $10^{-3}$
were selected. It corresponds to a negative deviation of about  
$7\sigma$ in the deviation distribution, 
depending on the total number of time bins in that
search. 
In total, 107 dip events were identified in 564-ks data, 
many of which appear more than once in searches 
with different bin sizes. 
The epochs of all the 107 events are listed in Table 1 for later referring and for readers'
convenience to find them in the RXTE data.
\begin{figure}
\epsfxsize=8cm
\epsffile{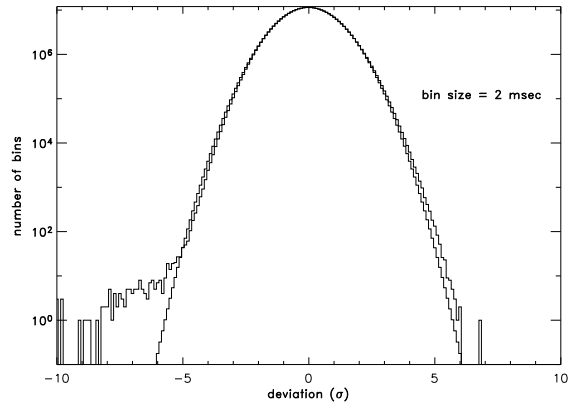}
\caption{
Example deviation distribution of the Sco X-1 RXTE/PCA light curve. 
See \citet{chang06} for the exact definition of the deviation.
The thick histogram is from the 564-ks RXTE/PCA data of Sco X-1 
and the thin one is a  Gaussian distribution, plotted for comparison. 
To show the relatively small number of events with large deviations, 
the ordinate is plotted in logarithmic scales. 
The excess at the negative deviation end 
is obvious. The small excess at the positive deviation 
part and the tiny deficiency between about $-3\sigma$
and $-5\sigma$ are due to the Poisson nature of the observed photons.  
}
\end{figure}

\begin{table*}
\label{eventlist}
 \centering
 \begin{minipage}{160mm}
  \caption{Epoch list of the 107 dip events.}
  \begin{tabular}{lll|lll|lll|ll}
  \hline
event & epoch (MJD) & & event & epoch (MJD) & & event & epoch (MJD) & & event & epoch (MJD) \\
\hline
 1 & 50127.47287408 & &	28 & 50560.64741831 & & 55 & 50820.82725554 & & 82 & 50999.28352289 \\ 
 2 & 50127.48015520 & & 29 & 50560.74484861 & & 56 & 50820.84952867 & & 83 & 50999.35461803 \\
 3 & 50227.88200130 & & 30 & 50560.79377724 & & 57 & 50820.89743563 & & 84 & 50999.41091596\\
 4 & 50227.95654701 & & 31 & 50560.80657724 & & 58 & 50820.97524034 & & 85 & 51184.53187584\\
 5 & 50228.08619841 & & 32 & 50561.59186087 & & 59 & 50820.98661576 & & 86 & 51186.48533043 \\
 6 & 50228.42335989 & & 33 & 50561.59809735 & & 60 & 50821.10690660 & & 87 & 51186.62284120 \\
 7 & 50228.94854367 & & 34 & 50561.60711487 & & 61 & 50821.50189352 & & 88 & 51187.09077119\\
 8 & 50229.20952234 & & 35 & 50561.60902924 & & 62 & 50821.92314378 & & 89 & 51187.81352256\\
 9 & 50229.28446027 & & 36 & 50562.67774632 & & 63 & 50963.22196721 & & 90 & 51187.95222912 \\
 10 & 50229.95584931 & & 37 & 50562.73583564 & & 64 & 50963.22696160 & & 91 & 51188.00380963 \\
 11 & 50229.95809071 & & 38 & 50664.04398279 & & 65 & 50963.29032203 & & 92 & 51188.01842742  \\
 12 & 50230.09181139 & & 39 & 50682.18029159 & & 66 & 50964.27470971 & & 93 & 51188.02324544\\
 13 & 50230.13947224 & & 40 & 50816.96028303 & & 67 & 50964.28074489 & & 94 & 51188.35803894 \\
 14 & 50230.95055234 & & 41 & 50816.96908550 & & 68 & 50964.28618818 & & 95 & 51188.61416433\\
 15 & 50231.07426689 & & 42 & 50817.04546707 & & 69 & 50964.29270126 & & 96 & 51191.87259403 \\
 16 & 50231.08555152 & & 43 & 50817.04865761 & & 70 & 50965.27044441 & & 97 & 51191.87461469\\
 17 & 50231.12912418 & & 44 & 50817.90240868 & & 71 & 50965.29753470 & & 98 & 51191.87770321 \\
 18 & 50522.78695518 & & 45 & 50818.05516266 & & 72 & 50965.34610105 & & 99 & 51191.93625665\\
 19 & 50522.89848183 & & 46 & 50818.77866986 & & 73 & 50966.12939997 & & 100 & 51192.94289686 \\
 20 & 50522.91066851 & & 47 & 50818.77931186 & & 74 & 50966.16083347 & & 101 & 51193.80435748\\
 21 & 50522.92443425 & & 48 & 50819.10373772 & & 75 & 50966.22742241 & & 102 & 51194.87184245 \\
 22 & 50522.95734852 & & 49 & 50819.62710164 & & 76 & 50996.34615057 & & 103 & 52348.09450448 \\
 23 & 50522.95752945 & & 50 & 50819.98466883 & & 77 & 50997.26933552 & & 104 & 52409.88829219\\
 24 & 50558.87343683 & & 51 & 50820.11086494 & & 78 & 50997.41920381 & & 105 & 52410.89054165\\
 25 & 50558.92531729 & & 52 & 50820.62648648 & & 79 & 50998.33941018 & & 106 & 52410.95092608 \\
 26 & 50559.86445702 & & 53 & 50820.70592722 & & 80 & 50998.34211168 & & 107 & 52413.98904378\\
 27 & 50559.93855131 & & 54 & 50820.71036958 & & 81 & 50999.27356237 & &  \\
\hline
\end{tabular}
\end{minipage}
\end{table*}

\begin{figure*}
\epsfxsize=17cm
\epsffile{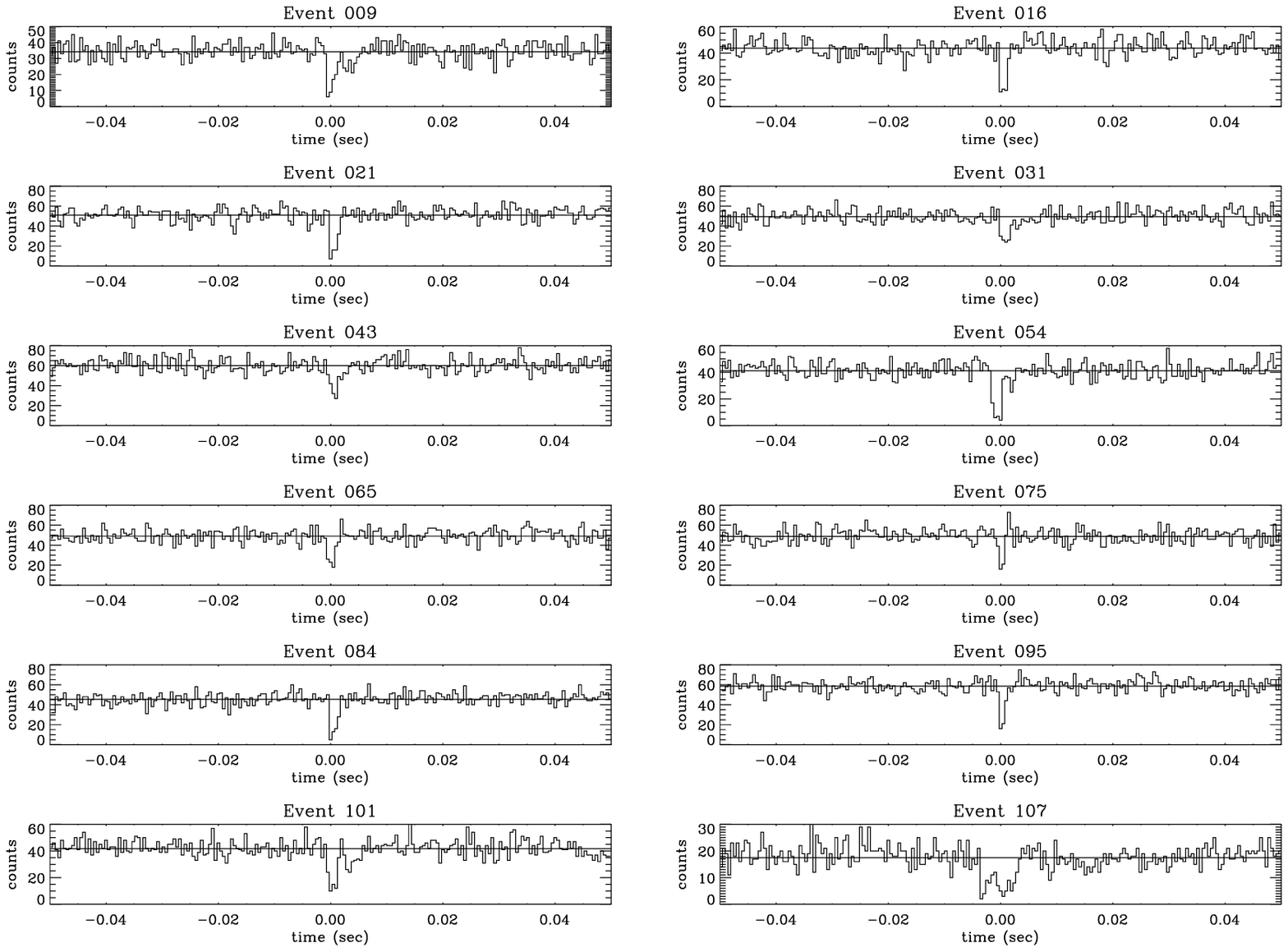}
\caption{
RXTE/PCA light curves of some example dip events. 
Horizontal lines are the average counts in each panel. 
The histograms are plotted with a bin size of 
0.5-ms. events are numbered in sequence according to their occurrence epoch. 
Different count rates for different observations are due 
to the intrinsic flux variation of Sco X-1 
and the number of PCUs turned on in that observation. 
}
\end{figure*}
Some example light curves of these dip events are shown here in Fig. 2. 
These light curves show a variety of characteristics.
About one half of all the events have a duration of 2 ms, 
and most of the rest are 1.5-ms and 2.5-ms events.
The longest one is 7 ms. The flux-drop fraction of these events, 
$A$,
defined as $A=(1-f)$ with $f$ being the ratio of the average counts
per bin within the event to that outside the event in a 100-ms window, 
is distributed between 0.3 and 0.8.
Fig. 3 shows the distribution of the 107 events in duration and in the flux-drop fraction.
The event duration is defined as the length of the time interval in which
the photon count number in a 0.5-ms bin is smaller
 than one standard deviation below the
average counts per 0.5-ms bin in a 100-ms window.
This definition, instead of considering all the bins with photon counts
just below the average, avoids a probably inadequate designation of a longer
duration to cases like event 31 in Fig. 2 and event 5 in Fig. 12.
However, a more complicated dip light curve may simply be missed with our definition, 
such as event 9 and 101 in Fig. 2. Adopting the average counts as the threshold
to define the duration will include the trailing dip of event 9 but not event 101.
To have an operational definition without arbitrary, subjective selection judgment,
we adopt our current definition.
This definition tends to somewhat
underestimate the duration and overestimate the flux-drop
fraction.
We note that our detection method as described 
above is biased against events of a small flux-drop fraction.
In the case of a search with time bins of 2 ms, for example, 
those time bins falling between $-5\sigma$ and $-7\sigma$ contain a mixture of random and good events
and are not selected (see Fig. 1).
This procedure is also biased against events much shorter than about 1 ms 
because of the small count number in one bin.
\begin{figure*}
\epsfxsize=17cm
\epsffile{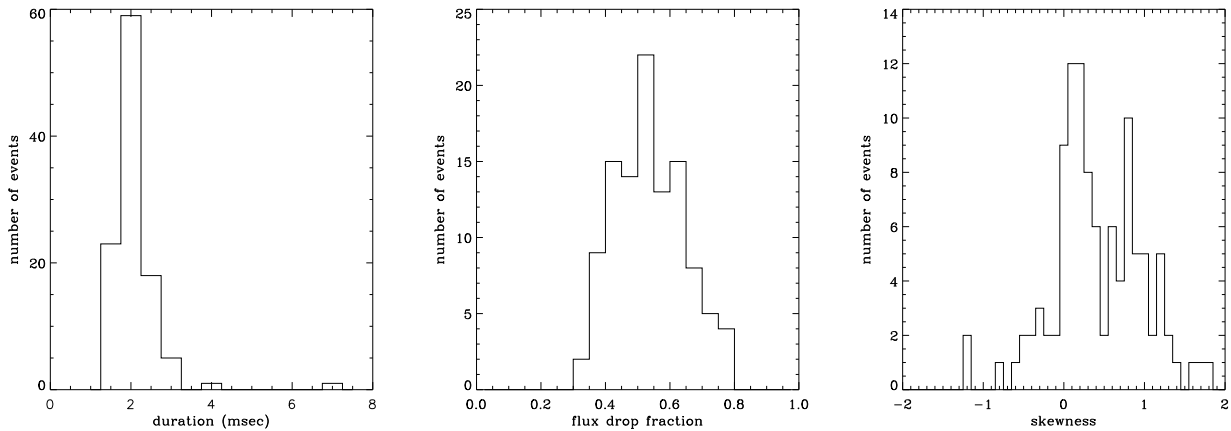}
\caption{
Distributions in the duration, flux drop fraction, and skewness  
of the 107 events.
}
\end{figure*}

Many of the events have an asymmetric light curve with fast drop and slow recovery,
like event 21 and 84 shown in Fig. 2. 
To characterize the degree of asymmetry, 
we define the 'skewness', $S$, of each event as
\begin{equation}
S=\frac{1}{N}\sum_x C_x\left(\frac{x-\langle x\rangle}{\sigma}\right)^3
\,\, ,
\end{equation}
where $N$ is the total number of `missing photons' 
within the event duration relative to the average count level
of a 100-ms window, $x$ is time and runs for each time bin, 
$C_x$ is the number of missing photons in time bin $x$,
and $\sigma$ is the standard deviation of this missing-photon
distribution in time ($\sigma^2=\sum_x C_x(x-\langle x\rangle)^2/(N-1)$ and 
$\langle x\rangle=\sum_x xC_x/N$). With such a definition, $N$ and $C_x$ are in general not integers. 
The skewness is then normalized to $\sqrt{15/N}$, which is roughly equal to the standard deviation 
in skewness randomly drawn from a Gaussian distribution
of $N$ photons \citep{press92}.
The skewness is zero for symmetric light curves and positive for light curves with fast drop and slow recovery.
The skewness distribution of the 107 events is shown in Fig. 3. 
It appears that there seems to be two populations of events in terms of skewness. 
%

\section{Signatures of possible instrumental effects}

It was pointed out in \citet{jones06} that many of these dip events may be
related to the so-called
`very large event' (VLE), in the RXTE terminology. 
Unfortunately, the VLE counts were recorded with a 125-ms resolution.
Besides, not all the individual VLE bins covering the dip epochs have an unusual high count number.
The association of VLEs and dip events is so far not conclusive.
Prompted by the report in \citet{jones06}, we examined 
some more particle-related properties of these dips,
in the hope to properly assess the instrumental signatures and to
identify non-instrumental dip events, if the current data allows. 

A VLE is an event that deposits more than about 100 keV energy in any one of
the six active xenon-layer anodes or the propane-layer anode 
in each Proportional Counter Unit (PCU) 
of the PCA on board RXTE.
Technical details of RXTE instrumentation can be found in the RXTE web site and also
in \citet{jahoda06}.
A VLE saturates the preamplifier and causes ringing in the signal chain 
when the amplified signal
is restored to the baseline.
A larger dead-time window is set for each VLE. For most of the Sco X-1 observations
it was chosen to be about 70 $\mu$s \citep{jahoda06}. 
The actual duration in which the signal chains are affected depends on the pulse height
of the saturating event.
The VLE counts are recorded with a 125-ms time resolution in the `standard-1' data mode
for every PCA observations.
The standard-1 data mode also records, with the same 125-ms resolution, the
propane-layer counts, which are events that trigger only the propane layer,
and the `remaining' counts, which are all the events except for `good' events,
VLEs, and propane events.
Good events are those trigger only one xenon anode. 
The propane layer is designed to distinguish events caused by soft electrons 
from those by X-rays. The `remaining' counts contain many `multiple-anode' events, 
which are usually treated as particle background.
Because Sco X-1 is very bright, most of the propane-layer events and the `remaining' 
events are in fact caused by photons from Sco X-1.
In most of the Sco X-1 observations, the data mode recording events triggering both
the `left' and `right' xenon anodes of the first layer (L1\&R1) 
with a 250-$\mu$s time resolution is also employed. 
Almost all of these L1\&R1 coincidence events are caused by photons from Sco X-1, 
instead of particles.
All the above information is useful for investigating the possible instrumental
effects for the dip events. Unfortunately the 125-ms resolution is much larger than
the dip event duration. More detailed information is recorded in the `standard-2' data
mode but with an even coarser resolution of 16 s.   

\begin{figure*}
\epsfxsize=17cm
\epsffile{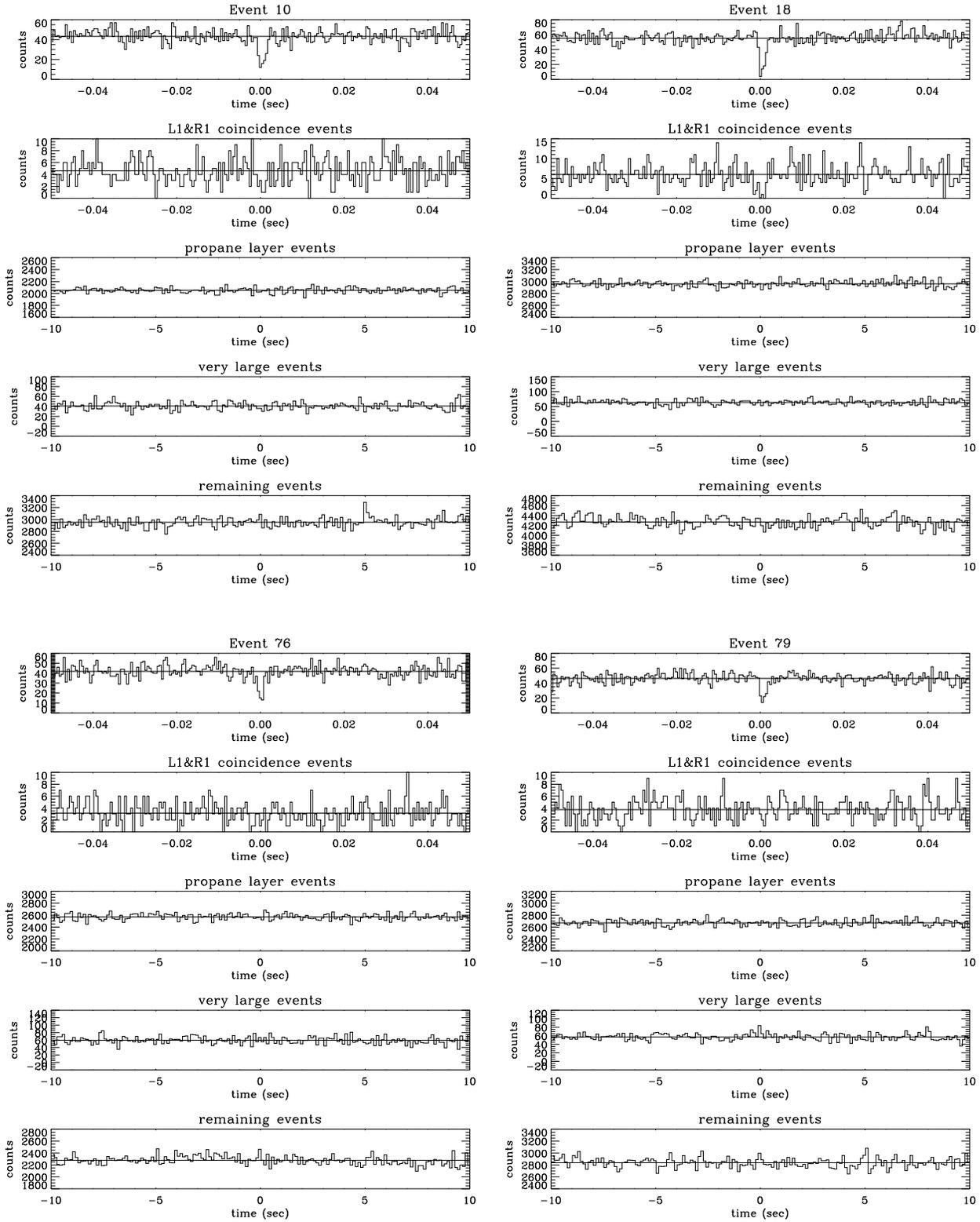}
\caption{
Example light curves of good counts, 
L1\&R1 coincidence counts, propane-layer counts, 
VLE counts, and `remaining' counts of 4 dip events. 
For each dip event, the time zero of all the 5 light curves 
is referred to the same epoch.
Please note the different time scales in the abscissae.
Each time bin in the light curve of good counts and L1\&R1 counts
is 0.5 ms. That in the other three is 125 ms.
None of these 4 is selected as a probable non-instrumental event; see
discussion in the text. 
}
\end{figure*}
\begin{figure*}
\epsfxsize=17cm
\epsffile{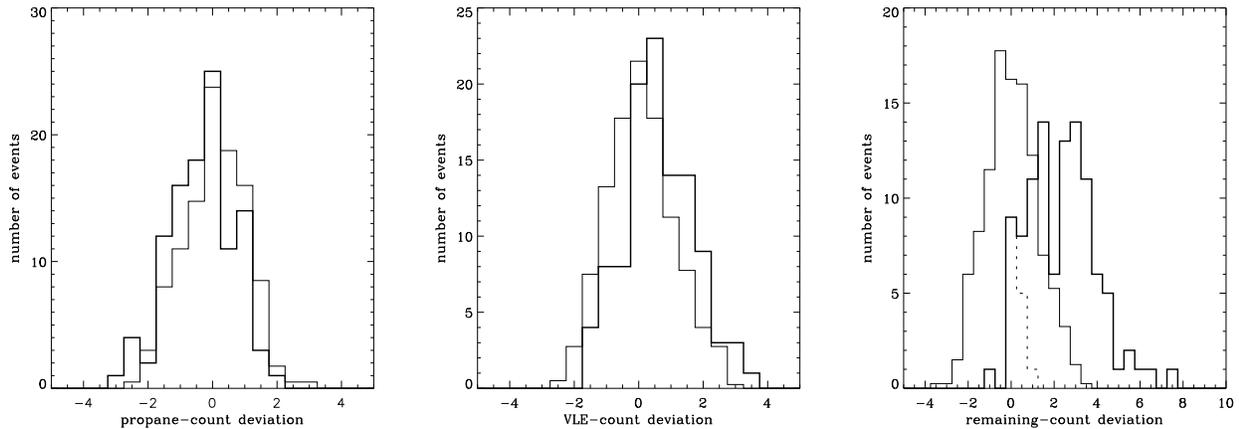}
\caption{
Deviation distributions of the propane counts, VLE counts and
`remaining' counts.
Thick histograms are for the 107 dip events and
thin ones are the average of 4 comparison groups.
The dashed line in the right panel is explained in Section \ref{noins}.
The `remaining' deviations of event 7, 101 and 107 are larger than 10
and are not included in the right panel. 
}
\end{figure*}
\begin{figure}
\epsfxsize=8cm
\epsffile{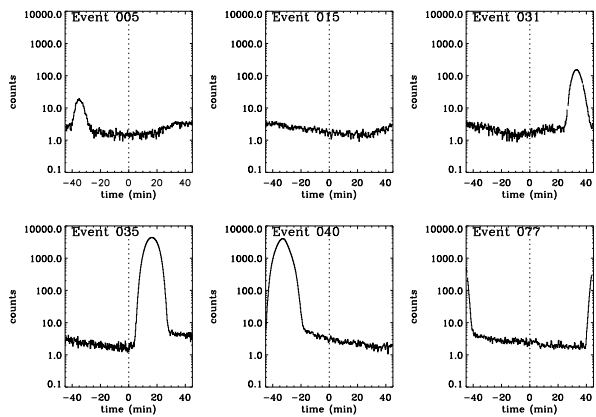}
\caption{
Some example light curves of HEXTE/PM counts.
The vertical dashed lines indicate the dip event epochs.
SAA passage is clearly seen in the HEXTE/PM counts.
Many of the 107 dip events are far away from the SAA passage, like
event 15 shown here.
The time difference  between the epochs of the dip and 
the nearest SAA peak passage is defined as the time since SAA. 
}
\end{figure}
\begin{figure}
\epsfxsize=8cm
\epsffile{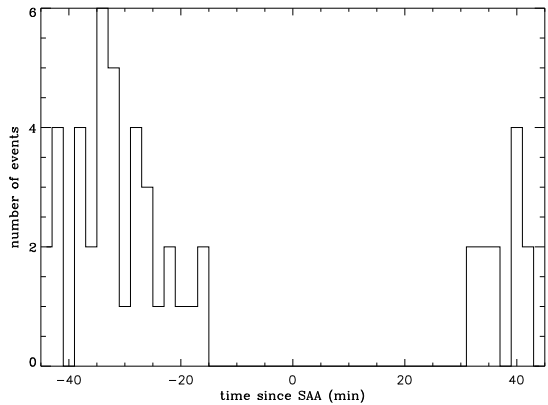}
\caption{
The distribution of dip events according to their `time
since SAA', defined in the caption of Fig. 6.
The histogram bin size is 2 minutes.
50 out of the 107 dip events are within 45 minutes before and after the SAA peak passage
and are all included in this figure. 
The gap around the SAA passage is because of data screening. 
}
\end{figure}
In Fig. 4, example light curves of good counts, 
L1\&R1 coincidence events, propane-layer events, 
very large events, and `remaining' events are shown. 
The time zero of all the 5 light curves is referred to the same epoch.
For all the 107 events, the L1\&R1 coincidence counts all drop consistently 
with the dip, although sometimes
not so obviously because of the small count number.  
The propane-layer events in fact do not show any anomaly. 
Many of the 107 events are associated with higher VLE counts, like event 79 in Fig. 4.
Their VLE counts are not unusually high. Only when all the events are studied together, 
the tendency of having a higher VLE count number becomes clear.
We note that, however, a higher VLE count does not always come with a dip event.
In Fig. 4, there are several VLE bins with a higher count number similar to that at the
dip epoch of event 79. There are no dip events in those time bins.
A clearer indication of possible instrumental effects appears in the `remaining' counts. 
Almost all the 107 events are associated 
with a `remaining' count number larger than the average in a
local window. 

To better present all these count excesses, 
we define, for each dip event, the propane deviation,
VLE deviation, and `remaining' deviation as the difference 
between the corresponding count number 
and the average in a certain window encompassing the bin in question.
The deviation is then further expressed in units of the square root of that average. 
These deviation distributions are plotted in Fig. 5. 
The average `remaining' count rate sometimes varies noticeably at time scales 
about 2 s or so. We therefore used a window containing 9 bins (about 1 s) for  
the computation of these deviations.
A comparioson deviation distribution is also shown together in Fig. 5.
The comparison distribution is the average 
of four distributions drawn from the bin at 
-4s, -2s, +2s and +4s away from the bin covering the dip epoch respectively. 
This comparison distribution
is a representation of the distribution without contamination 
from the suspected particle events.  
One can see that there is nothing unusual in the propane-count deviation.
The VLE deviation distribution exhibits a positive shift but still contains a considerable
population of negative deviation. 
The `remaining'-count deviation distribution shows a very significant positive shift.

\begin{figure*}
\epsfxsize=17cm
\epsffile{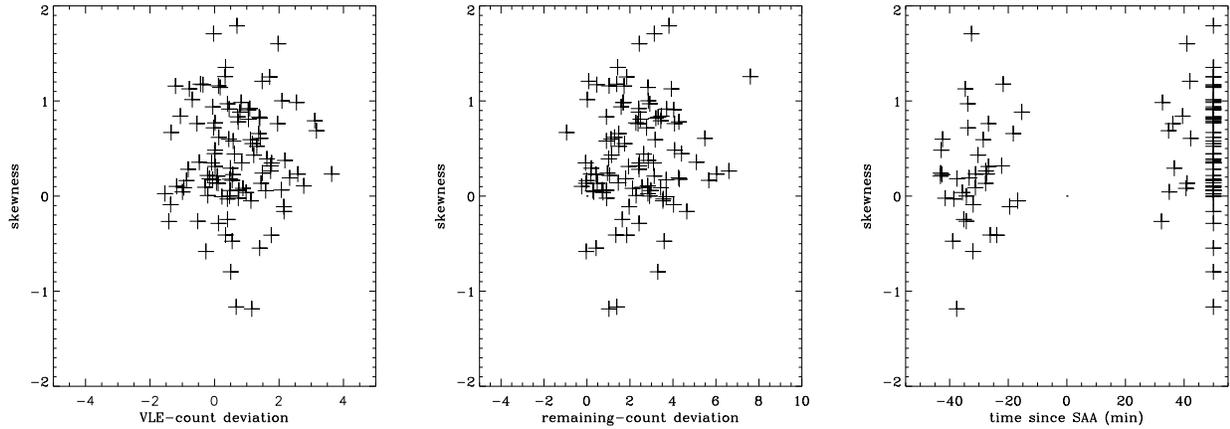}
\caption{
The VLE deviation, `remaining' deviation and time since SAA
versus the dip skewness.
Event 7, 101 and 107, with a `remaining' deviation larger than 10,
 are not included in the central panel.
In the right panel, all the dips outside the $\pm 45$-min window of an
SAA peak passage are plotted at the time since SAA equal to
50 min. 
}
\end{figure*}
To further investigate the possible particle association, 
we computed the time difference between
the dip epoch and the nearest SAA peak passage. 
It was done with the data obtained with the two particle monitors
of the instrument HEXTE \citep{rothschild98}. Some examples are shown in Fig. 6.
The event distribution in the time since SAA is shown in Fig. 7, in which
all the dip events within 45 minutes before and after the SAA peak passage are included.
One can see that there is no indication of any correlation 
between the dip epoch and the SAA passage.   
The suspected particle events are not related to the SAA. 
We have also examined other characteristics of these dip events and
found no obvious correlations among the duration, 
flux-drop fraction, skewness, VLE deviation,
`remaining' deviation and time since SAA either. 
That the dip skewness is not related to the VLE or
`remaining' deviations is somewhat puzzling. 
The skewness versus those deviations and the time since SAA
is plotted in Fig. 8. 

68 dip events out of the 107 are in observations with 
a data mode carrying two-channel spectral information.
The energy bands for these two channels are about 2-5 keV and 5-18 keV. 
There seems to be a trait that the flux drop
fraction is higher in the high energy band than in the 
low energy one. Some examples are shown in Fig. 9.
The distribution of the ratio of the flux-drop fraction in the high energy 
band to that in the low energy one for each dip events
is plotted in Fig. 10. Occultation events, because of diffraction, 
will have a smaller flux drop in a lower energy band, while the instrumental 
dead-time effect does not have that 
energy dependence. Although random fluctuation may render 
individual identification unreliable,
Fig. 10 suggests that, statistically speaking, some dip events 
may not be caused by instrumental effects. 

It was proposed in \citet{jones06} that 
`the dips are caused by electronic dead time in response 
to some type of charged particle shower in the spacecraft.'
Indeed, the positive deviation in the VLE
and `remaining' counts associated with dip events within 125-ms uncertainty
(Fig. 5) indicates that, at least statistically speaking,
many of the dip events are likely due to the dead-time effect caused by 
particles.
To have such a large flux drop for such a long period, however, 
several PCUs must be affected 
at the same time and a very large amount of charged particles is needed.
The PCA has 5 almost identical PCUs. The usual signal dead time is about 10$\mu$s,
and that for a VLE signal is about 70$\mu$s. 
That the `remaining' excess is much more significant than the VLE one indicates
that the dead time caused by the hypothesized lower-energy secondary particles
or by the VLE ringing events
may be the major, direct cause for these dips.
Although the Crab nebula is about 8 times dimmer than Sco X-1 and therefore individual
dips cannot be identified, an excess at the negative end of the deviation distribution
may be expected if the data amount is enough.
However, because the flux-drop fraction of these dips is not close to unity, but 
distributed between
0.3 and 0.8, 
they, at a rate of about one per 5 ks, are not noticeable in the deviation distribution drawn 
from the 380-ks RXTE/PCA archival data of the Crab nebula. 
On the other hand, the indication so far is not strong enough
to argue that all the dips are caused by the cosmic-ray induced dead-time effect.
There are dip events whose associated `remaining' and VLE deviations are negative or very small.
There are time bins (125 ms) with large VLE or `remaining' excesses but without any dips.
Furthermore, the flux-drop fraction of the dip events is generally larger in the high-energy
band than in the low-energy one, a signature of diffraction. 
It is likely that these dips are heavily contaminated by particle-related events, 
but not all of them are caused by the instrumental dead-time effect.

\section{Events without indication of instrumetal effects}
\label{noins}

Limited by the 125-ms time resolution of the standard-1 data, it is 
impossible to definitely identify individual dip events as caused by
the instrumental dead-time effect or as of a probable astronomical origin.
Nevertheless, as discussed in the previous section, it is likely that
these dip events are a mixture of the two kinds. 
Since the light-curve skewness and the `time since SAA' are not correlated with
the VLE or `remaining' deviations,
we propose to select those probable non-instrumental
dip events solely based on the two deviations. 
From Fig. 5, one can see that the comparison population has a half width at half maximum
of about 1.25 in both deviation distributions. If a non-instrumental population exists, 
the half width
in the deviation distribution is about the same. 
In this regard, a somewhat conservative criterion to identify
probable non-instrumental events can be set as
the VLE and `remaining' deviations being both smaller than 1.25. 

From the `remaining' deviation distribution in Fig. 5, the total number of the 
probable non-instrumental events can be estimated 
by assuming all the events with a negative `remaining' deviation
are the probable non-instrumental events and they have 
a more or less symmetric deviation distribution centered at zero deviation.
Such a possible distribution is plotted in dashed lines in Fig. 5.
The total number estimated this way is about 16. This number is of course 
somewhat arbitrary and only suggestive.
The sum of the two deviations can be taken as a measure 
and the second criterion can be set,
to be conservative again, as that sum being smaller than 1.0 
so that the total number is 12.
These two criteria are also shown in Fig. 11. 
\begin{figure*}
\epsfxsize=17cm
\epsffile{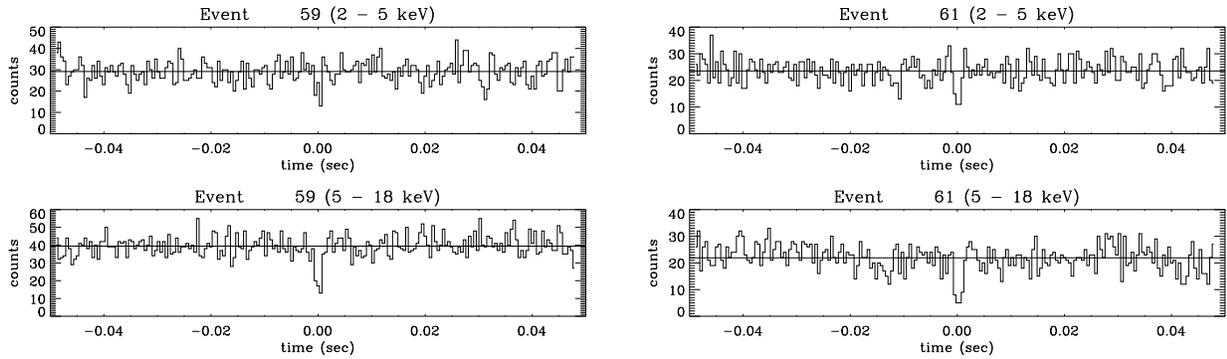}
\caption{
The light curves of event 59 and 61 in two energy bands.
}
\end{figure*}
\begin{figure}
\epsfxsize=8cm
\epsffile{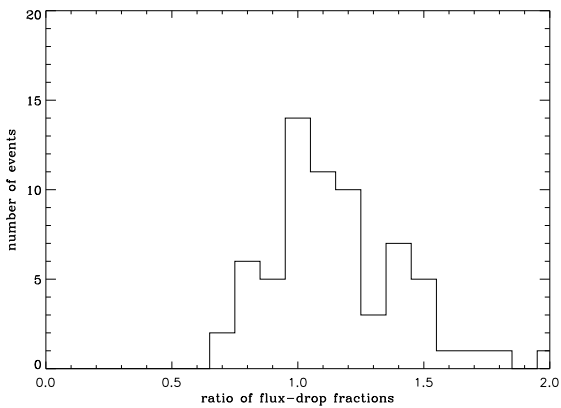}
\caption{
Distribution of the flux-drop fraction ratio. 
The ratio is the flux-drop fraction in the high-energy band (5-18 keV)
divided by that in the low-energy one (2-5 keV).  
}
\end{figure}
\begin{figure}
\epsfxsize=8cm
\epsffile{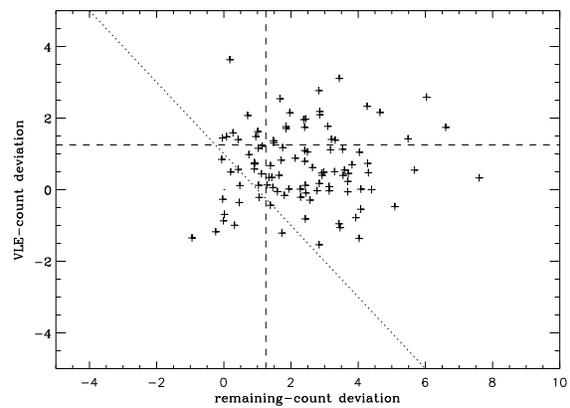}
\caption{
VLE-count deviation versus `remaining'-count deviation for all the
dip events excluding event 7, 101 and 107, 
which have a `remaining' deviation larger than 10. 
The dashed lines are discussed in the text.
}
\end{figure}
The light curves of the 12 probable non-instrumental events 
are shown in Fig. 12.

The second criterion is somewhat arbitrary. 
We have conducted such a selection procedure with different
window size for computing the `remaining' deviation. 
8 dip events among the 12 are always selected even when
the window is as large as 10 s. The 4 left out (event 41, 52, 59, and 74)
are all those with the `remaining' count level varying at 2-s time scales.
We think the 1-s window, which we employed to compute the `remaining' 
deviation distribution shown in
Fig. 5, is a more meaningful one and therefore adopt the 12 dip events selected 
with the procedure mentioned above
as the probable non-instrumental events. 
\begin{figure*}
\epsfxsize=17cm
\epsffile{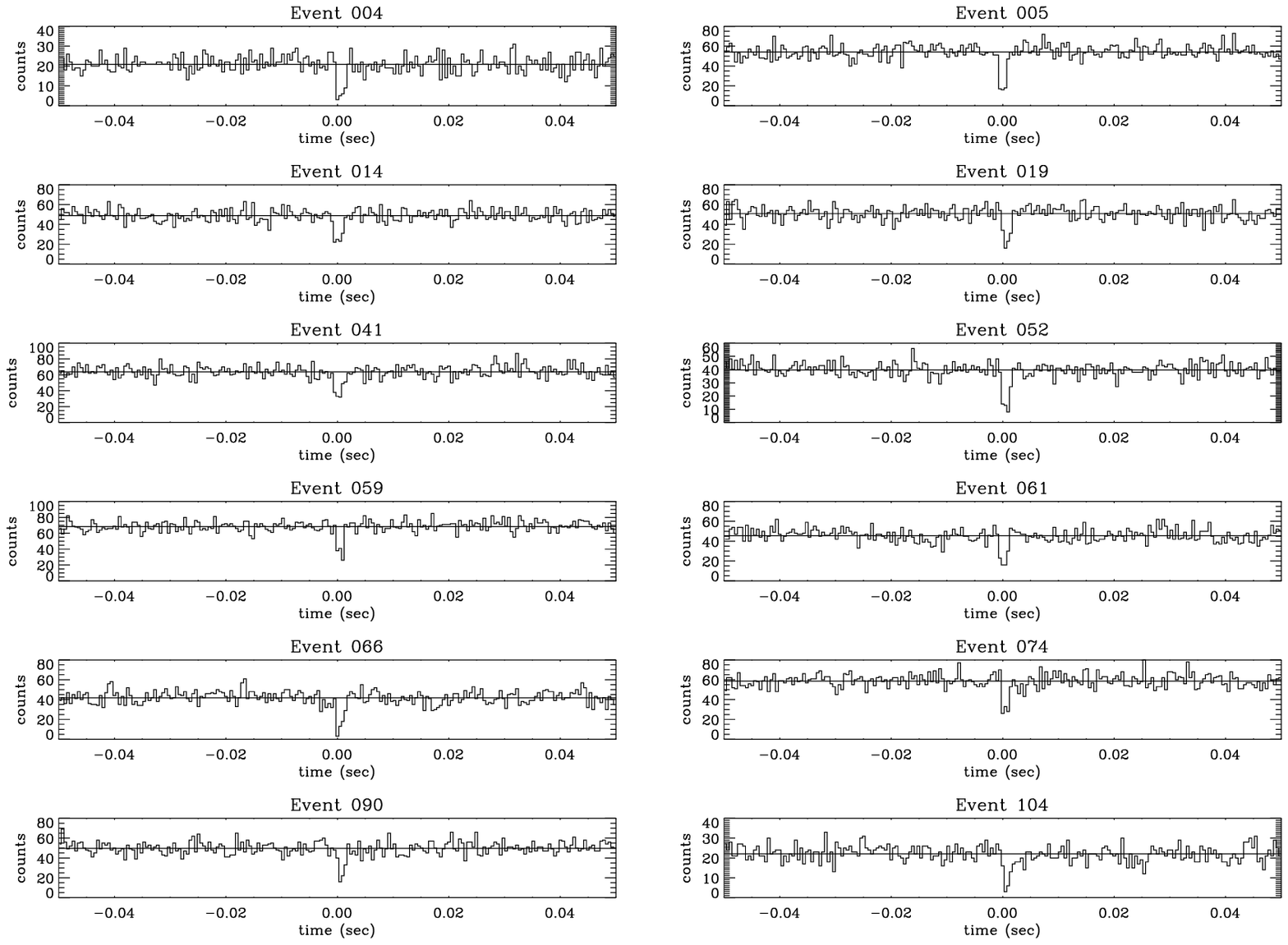}
\caption{
Light curves of the 12 proposed probable non-instrumental events.
}
\end{figure*}
Among the 12 probable events, 7 (event 19, 41, 52, 59, 61, 66, and 74) have two-channel
spectral information. 
Their flux-drop-fraction ratios of high- to low-energy bands are all
larger than unity.
Only event 4 and event 66 show obvious asymmetry in their light curve, 
that is, with fast drop and slow recovery in the flux.
We suggest that, in the context of the above discussion, 
the 12 events, whose light curves are plotted in Fig. 12,
are probable non-instrumental events.
It is about 10\% of all the identified dip events.

\section{TNO occultation and size distribution} 

Hereafter we assume the 12 probable events proposed above are really non-instrumental.
These events of abrupt and significant flux drop at millisecond time scales 
are not random fluctuations, as evidently
shown in the deviation distribution (e.g.\ Fig. 1). 
Based on spectral analysis, X-ray emission from Sco X-1 is believed to originate 
from the neutron star surface thermal emission
and corona comptonization near the inner accretion disk 
(e.g.\ \citet{bradshaw03,barnard03}).
A sudden quench of X-ray emission from Sco X-1 in such a short time scale 
seems very unlikely.
We have also examined the X-ray color for those events with two spectral channels. 
No detectable color change in the X-ray emission before and after any event was found.  
Occultation by objects in the line of sight is the most viable cause for these events.

The flux-drop fraction, $A$, of these events ranges from 0.36 to 0.72. 
To have such a significant flux drop, 
the occulting body must be a few times larger than the corresponding Fresnel scale, 
which is $\sqrt{\lambda d/2}$, where $\lambda$
is the wavelength and $d$ is the distance. 
It is about 30 m for $d=40$ AU and $\lambda=0.3$ nm 
(4 keV; most photons in the data are at this energy).
\begin{figure*}
\epsfxsize=17cm
\epsffile{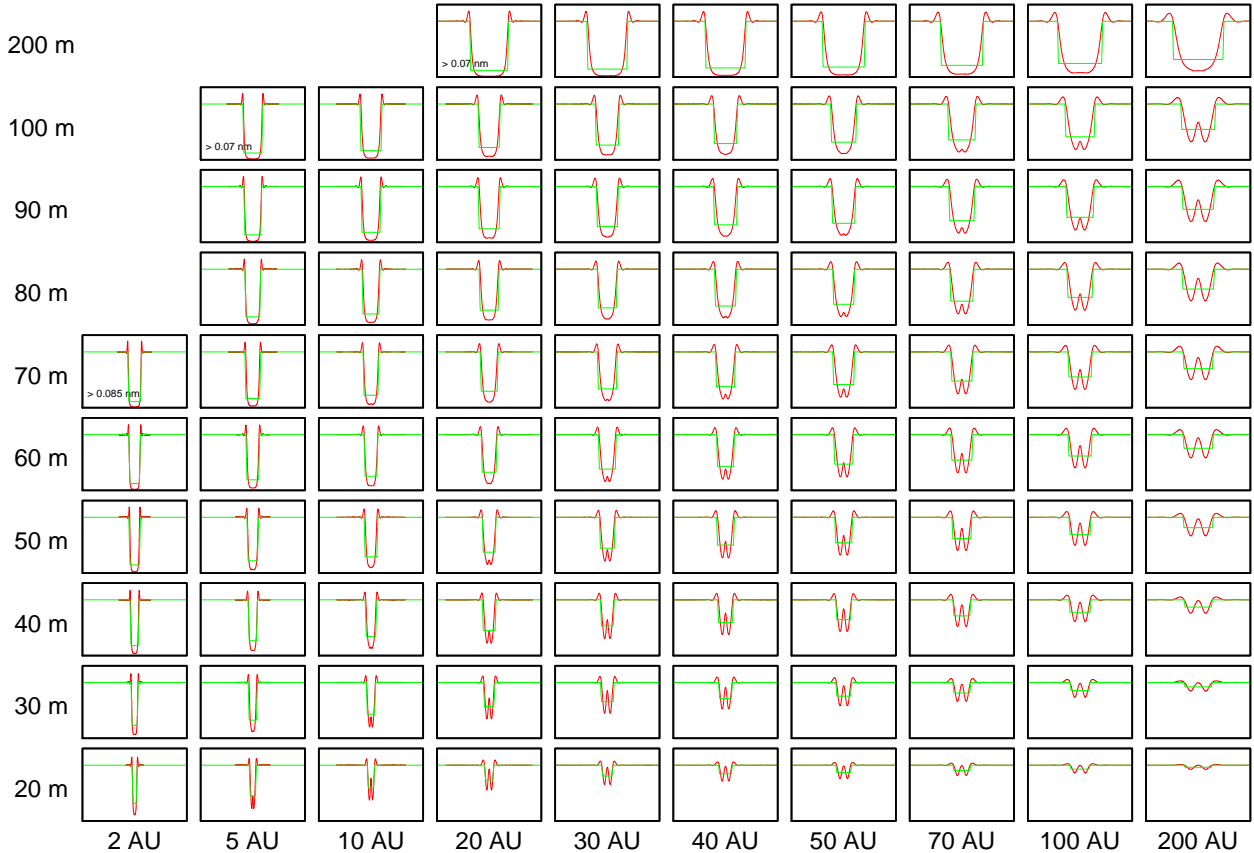}
\caption{
X-ray diffraction patterns (one-dimensional cut through a shadow) for objects of
different sizes at different distances.
The X-ray spectrum considered here is that of the RXTE/PCA observed
photons from the Sco X-1, 
roughly in the energy range from 2 keV to 20 keV and peaked at
about 4 keV. The effect of the varying spectrum of Sco X-1 on these patterns is
negligible. The occulting object is assumed to be spherical. The patterns shown here
assume a crossing impact parameter of 0.33, which is the distance from the shadow center in
units of the object size.
The width of each panel is 600 m.
An equivalent square function pattern is also
plotted in each panel, whose depth is the flux-drop fraction $A$
discussed in the text.
One can see that to have a flux-drop fraction of 0.3, the object is about 
the corresponding Fresnel scale of 4-keV photons. 
Note that the width of the square function here is wider than the corresponding 
duration of the dip events as we define in this paper. 
}
\end{figure*}
Some X-ray diffraction patterns for objects of different sizes 
at different distances are
computed and plotted in Fig. 13, with the PCA-detected photon 
spectrum from
Sco X-1 as the input radiation for the computation.
Fig. 13 shows the case for a spherical blocking object and 
a crossing impact parameter equal to 0.33.
The crossing impact parameter is the distance to the shadow center in units of the object size.
Computations with different crossing impact parameters, such as 0 and 0.67,
 do not yield too different
a flux-drop fraction, although the diffraction patterns are different, 
particularly regarding the central brightening in the shadow. 
From the diffraction pattern computation, one can see that to have $A\approx 0.3$, 
the size of the occulting 
body at 40 AU away must be larger than 30 m, the Fresnel scale of 4-keV photons. 
We note that, 
given a certain ratio of the object size to the square root of the distance,
the diffraction patterns are the same, up to a scaling factor in length 
proportional to the object size
or the square root of the distance
(e.g.\ \citet{roques87}). 
The flux-drop
fraction $A$ is therefore also the same for these patterns. 
At the distance of Sco X-1, 2.8 kpc \citep{bradshaw99}, 
the Fresnel scale is about 100 km (hereafter all the Fresnel scales
are referred to that for 4-keV photons).
It is clear if these events are occultation by objects moving around Sco X-1, 
those objects must be larger than about 100 km 
and also larger than the X-ray emitting region, which ranges from
a few hundred kilometers to about 50,000 km, depending on different models 
(e.g.\ \citet{bradshaw03,barnard03}). 
To produce a 2-ms event, the speed of those objects needs to be larger than $5\times 10^4$ km/sec, 
a very unlikely speed. If we consider, more realistically, 
the event duration as the shadow size instead of the object size, an even higher speed  
is required.
The same argument can be applied to other objects in the interstellar space. 

For objects at the distance of the Oort Cloud, typically 10,000 AU, 
the Fresnel scale is about 500 m. 
Again, to produce the events that we found, 
the Oort Cloud objects need to be larger than that scale. 
It is unlikely that a kilometer-size object in the Oort Cloud 
can produce millisecond events, 
since its speed relative
to the earth is only about 30 km/sec. 
On the other hand, other solar system small bodies 
at a shorter distance can satisfy the duration
and flux-drop constraints at the same time. 
A major source candidate is the main-belt asteroids. 
Their expected event rate for occulting
Sco X-1, however, is too low ($10^{-9}$ sec$^{-1}$, down to 10-m size; 
see \citet{chang06}). 
We therefore propose that the 12 events, if really not instrumental,
are occultation caused by objects in the trans-Neptunian region, that is, by TNOs.

To pin down the distance and the size of
each occulting body more accurately for all the events is an important issue
but requires further analysis on the dead-time-corrected light curves, 
diffraction patterns and different orbital inclinations and eccentricities. 
That study will be reported in another forthcoming paper. 
For an approximate estimation of
the occulting-object sizes at this stage, 
we contend ourselves by considering a typical relative speed of 30 km/s between the
occulting TNOs and RXTE and a random crossing through the shadow.
The average crossing length of a random crossing is $\frac{\pi}{4}$ times that
of a central crossing, assuming a circular shadow.
We therefore set the size range of these 12 objects to be
from 60 m to 100 m for durations from 1.5 ms to 2.5 ms. 
We note that
the orbital speed of the Earth is about 30 km/s, RXTE's speed relative to the Earth
is about 7 km/s, a TNO at 43 AU has an orbital speed about 5 km/s 
for a low eccentricity orbit, the shadow size is larger than the object, and 
our definition of the event duration tends to somewhat underestimate.

We next proceed to estimate the total number of TNOs of these sizes.
For a background point source, the event rate is
\begin{equation}
\frac{N}{T}=\frac{\int_{s_1}^{s_2}\left(\frac{{\rm d}N}{{\rm d}s}\right)
sv\,{\rm d}s}{d^2\Omega_{\rm A}} \,\,\,\, ,
\end{equation}
where $N$ is the number of detected events 
(assuming a 100\% detection efficiency), $T$ the total exposure time,
$\left(\frac{{\rm d}N}{{\rm d}s}\right)$ the differential size distribution,
$v$ the typical sky-projection relative speed, $d$ the typical distance to the TNOs,
 and $\Omega_{\rm A}$ the total solid angle of the sky distributed with TNOs.
To derive  
$\left(\frac{{\rm d}N}{{\rm d}s}\right)$ from the event rate, a functional
form of the distribution needs to be assumed. 
On the other hand, if the integration is only over a small range of the size,
we may derive an average value at that size.
Noting that 
$\frac{{\rm d} N}{{\rm d}\log s}=\frac{{\rm d} N}{{\rm d}s}\,s\,\ln 10$,
we have
\begin{equation}
\int_{s_1}^{s_2}\left(\frac{{\rm d}N}{{\rm d}s}\right)
sv\,{\rm d}s
=
\left(\frac{{\rm d}N}{{\rm d}\log s}\right)_{s_1<s<s_2}\frac{v(s_2-s_1)}{\ln 10}
\,\,\, ,
\end{equation}
and
\begin{equation}
\left(\frac{{\rm d}N}{{\rm d}\log s}\right)_{s_1<s<s_2}
=
\frac{d^2\Omega_{\rm A}}{v(s_2-s_1)}\frac{N}{T}\,\ln 10
\,\,\, .
\end{equation}

We can estimate $\Omega_{\rm A}$ with the inclination 
distribution obtained from the CFHT survey 
\citep{trujillo01}, which reported a 20$^\circ$ half-angle 
of an assumed gaussian distribution. 
This half-angle in the inclination distribution
translates to about 12.8$^\circ$ for a corresponding half-angle 
in the apparent ecliptic latitude distribution, assuming
circular orbits. Sco X-1 is 5.5$^\circ$ north of
the ecliptic. To use the detection rate in the direction 
toward Sco X-1 to represent the whole TNO population,
the equivalent sky area is a zone occupying $\pm 15.5^\circ$ in latitude.
Therefore, $\Omega_{\rm A}=360\times 31\times(\frac{\pi}{180})^2=3.4$.
Assuming a typical distance $d=43$ AU, a typical relative sky-projection speed
$v=30$ km/s, 
and $T=564.3$ ks, we have in the size range from 60 m to 100 m 
\begin{equation}
\left(\frac{{\rm d}N}{{\rm d}\log s}\right)
\approx 5.7\times 10^{15}
\,\,\, .
\end{equation}
The total number of TNOs in that size range is about $1.3\times 10^{15}$.

\begin{figure}
\epsfxsize=8cm
\epsffile{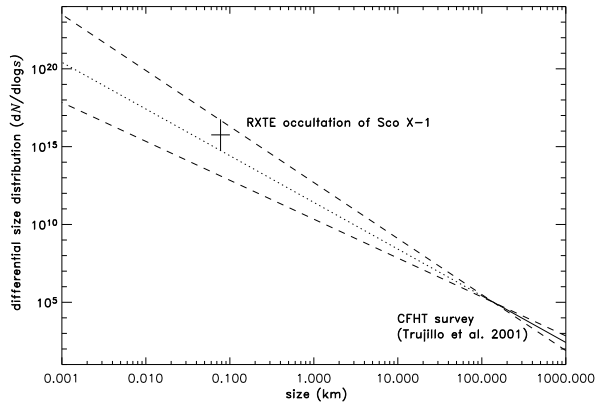}
\caption{
The differential size distribution of TNOs. 
The total number of TNOs per decade of size ($\frac{{\rm d}N}{{\rm d}\log s}$) inferred from the 
RXTE occultation of Sco X-1 is plotted as the cross with
a symbolic uncertainty level of a factor of ten. 
The solid and dotted lines are the best fit from the CFHT survey \citep{trujillo01}
and its extrapolation towards smaller sizes.
The dashed lines mark its 1-$\sigma$ uncertainty in the power index of the distribution.
Direct imaging of TNOs down to size of about 30 km 
were reportedly achieved by the Hubble Space Telescope \citep{bernstein04},
whose estimated TNO size distribution at about 30 km is a factor of 25 below the extrapolation
of the afore mentiond CFHT survey. 
The Taiwanese-American Occultation Survey (TAOS) 
is expected to detect occultation events caused by TNOs 
of kilometre size \citep{alcock03}. 
}
\end{figure}
There are considerable uncertainties in the above estimate.
First of all, the suggestion that 12 out of the 107 dip events 
are non-instrumental
is inconclusive. The detection efficiency of the procedure 
that we employed needs to be investigated in more details.
The size of these possible occulting TNOs is only very roughly estimated.
With all these cautions in mind, we tentatively discuss
the implication of the discovery of TNO X-ray occultations.     
The estimated total number, about one quadrillion, 
for TNOs of size between 60 m and 100 m is several orders
of magnitude larger than usual model predictions, 
but is close to, within the 1-$\sigma$ level, the direct extrapolation
of the distribution
obtained from the CFHT survey \citep{trujillo01}, which gives a best fit of 
${\rm d}N/{\rm d}s\propto s^{-q}$ with $q=4.0^{+0.6}_{-0.5}$ (1 $\sigma$) 
and the total number for TNOs 
larger than 100 km equal to $3.8^{+2.0}_{-1.5}\times 10^4$ (Fig. 14). 
Our result seems to indicate that the power-law size distribution 
of TNOs may extend down to the order of
100 m without any breaking or flattening.
On the other hand, 
if the assumed albedo is appropriate 
when converting the luminosity function into a size distribution
from the HST observation \citep{bernstein04},
and if that size distribution inferred from 
3 detections in 0.02 square degrees 
of the sky is reliable, 
a break will occur 
at about 50 km and there must exist a second component at smaller sizes. 
This second component
may require a new mechanism in the theory of planet formation.

\section{Summary}

We report 107 dip events of millisecond time scales in 564.3-ks RXTE/PCA archival data of Sco X-1. 
These dip events are apparently contaminated by those due to
instrumetal dead-time effects induced by high-energy cosmic rays.
We performed detailed analysis and propose that about 10\% of these dip events
are probable TNO-occultation events. The inferred size distribution of TNOs at size
about 100 m is several orders of magnitude larger than current theoretical predictions.
Confirming or disproving this result is important for
our understanding of planet formation processes in the early solar sytem.
X-ray diffraction patterns in the occultation light curves may also be exploited to 
study the background X-ray source, potentially in a very high angular resolution.
It is unfortunately impossible to definitely separate true occultation events from
instrumental ones with currently available data.
Observations of RXTE/PCA with an adequately designed data mode or of other new instruments
in the future
may resolve this issue.

\section*{Acknowledgments}

We thank R. Rothschild, F. Roques, and G. Georgevits for helpful discussion.
The examination of the event rate
versus the time-since-SAA with the SAA passage determined from HEXTE/PM data
were first done by Tommy Thompson and Richard Rothschild. 
This research has made use of data obtained through 
the High Energy Astrophysics Science Archive Research Center Online Service, 
provided by the NASA/Goddard Space Flight Center, 
and of the JPL HORIZONS on-line solar system data and ephemeris computation service. 
This work was supported by the National Science Council of 
the Republic of China under grant NSC 95-2112-M-007-050 and by Academia Sinica in Taipei.

\label{lastpage}
\end{document}